# Mobility and dynamics of Giant landslides

Julien Gargani[1,2]

[1]Univ. Paris-Saclay, CNRS, Geops, Orsay, France

[2]Univ. Paris-Saclay, Centre d'Alembert, Orsay, France

julien.gargani@universite-paris-saclay.fr

**Highlights**

- One or a few landslide events created the large scar and mass deposits in Tahiti.
- Peak slide velocity reached 125–250 m/s.
- The high mobility related to an effective basal friction of 0.2–0.3 under submarine conditions.

**Abstract**

The rarity of large landslides reduces the number of observations and hinders the understanding of these phenomena. Runout distance was used here to determine whether the large landslide deposit formed several thousand years ago in northern Tahiti was caused by a single or multiple events. Using modelling to quantify the dynamics of this event suggested that a single event or a small number of events (n<10) were responsible, and that the maximum slide velocity was high (>125 m/s) under partially submarine conditions. Such submarine propagation favoured a slower dynamic but a longer runout. The effective basal friction under submarine conditions ranged from $0.2 < \mu < 0.3$.

# 1. Introduction

Various hypotheses have been proposed to explain the triggering of large landslides: (i) ground acceleration due to earthquakes (Keefer, 1994) or volcanic eruptions (Carracedo et al., 1999), (ii) new stress triggered by dyke intrusion or magma reservoir deformation (Carracedo et al., 1999; Gargani et al., 2006; Hampel and Hetzel , 2008;  Le Corvec and Walter, 2009), (iii) increased pore pressure associated with higher rainfall, sea level or lake level (Muller-Salzburg, 1987;  Mc Mutry et al., 2001; Cervelli et al., 2002;  Kilburn and Petley, 2003; Veveakis et al., 2007; Quidelleur et al., 2008; Crozier, 2010; Gargani et al., 2014; Iverson et al., 2015), and (iv) slope change due to deep erosion or vertical movement. The complexity and inter-relatedness of the processes involved often prevent clear conclusions regarding the precise causes of slope failures.

The nature of large landslide dynamics along the slide path is still under debate because the mobility of these landslides is often larger than expected (Legros, 2002; Lucas and Mangeney, 2007; Iverson et al., 2015). It may be necessary to consider complex processes of various origins to explain such long-distance sliding, such as high basal pressure (Goren and Aharonov, 2007), ground vibration or hydroplaning (Hürlimann et al., 2000) under submarine conditions (Gargani and Rigollet, 2007; Gargani et al., 2008), presence of ice or evaporites (Bigot-Cornier and Montgomery, 2007; Crosta et al., 2018, Jouannic et al., 2012.; Jouannic et al., 2015), and submarine or aerial propagation (De Blasio, 2011b). The complexity of these processes also leads to difficulties in modelling landslides. For example, the friction angle of actual materials is often higher than that of materials used in simulations.

Several studies have proposed that deposit geometry could provide indications of landslide dynamics (Puga-Bernabéu et al., 2017; Crosta et al., 2018; Collins and Reid, 2020). For example, runout distance may relate to landslide dynamics and can be used to determine the processes involved (Crosta et al., 2018) as it may depend on the effective friction during sliding (Staron and

Lajeunesse, 2009). The spreading of landslide material may reflect the processes involved (Puga-Bernabéu et al., 2017; Collins and Reid, 2020). Although morphologic features represent a good set of observable data for past landslides, in the case of large landslides, it may be difficult to discriminate between the specific mobility of the material and the normal spreading of a large volume of material when using morphological data. In addition, large landslide dynamics are difficult to understand due to a lack of accurate data, as deformation and erosion may have affected their morphological characteristics following landslide occurrence (Menendez et al., 2008). Moreover, the rarity of these events reduces the chances of direct observation and thus hinders understanding; dedicated studies are primarily based on past events (Puga-Bernabéu et al., 2017). Large landslide morphology is often recorded only in part due to the difficulty of access, as most cases are located underwater (Abril and Periáñez, 2017), are covered by more recent sediments or lava flows, or are located on other planets like Mars (Quantin et al., 2004). Furthermore, landslide scar and mass transport deposit could represent the consequences of several events triggering the development of complex morphologies (Johnson and Campbell, 2017). Given these observational difficulties, it is not always easy to discriminate morphologies generated by a single landslide from those generated by multiple events.

Field studies, geochronological results, and bathymetric data suggest that a large mass displacement took place in Tahiti 872 ± 10 kyr ago (Clouard et al., 2001; Hildenbrand et al, 2004, 2006). However, the cause of the original slope destabilisation and the resulting landslide dynamics have never been precisely constrained. First, it is not clear whether this landslide scar and associated deposits are the consequence of a single large event, a few moderate events, or many small events. Second, the velocity of the Tahitian slide has never been quantified. Field investigations and seismic observations have constrained the geometry of past landslides but have not focused on quantitatively describing their dynamics, while modelling allows the quantitative assessment of slide velocity along the path. Landslide speed has often been estimated for smaller landslides (Rodriguez et al., 2017; Salmanidou et al., 2018), but data and simulations are less common for large landslides. Third, the influence of submarine versus aerial propagation on the velocity of large landslides is unclear, as the former results in increased drag force and decreased frictional force due to the incorporation of water into the slide; interactions between these factors can generate diverse dynamics.

Numerical experiments could provide a complementary approach to previous analyses of the Tahitian landslide scar and related deposits, potentially allowing the validation of a specific scenario and providing quantitative insights. Therefore, this study aimed to (i) discriminate between three hypotheses for this landslide's dynamics (single large event, few moderate events, or many small events) and (ii) estimate the maximum slide velocity.

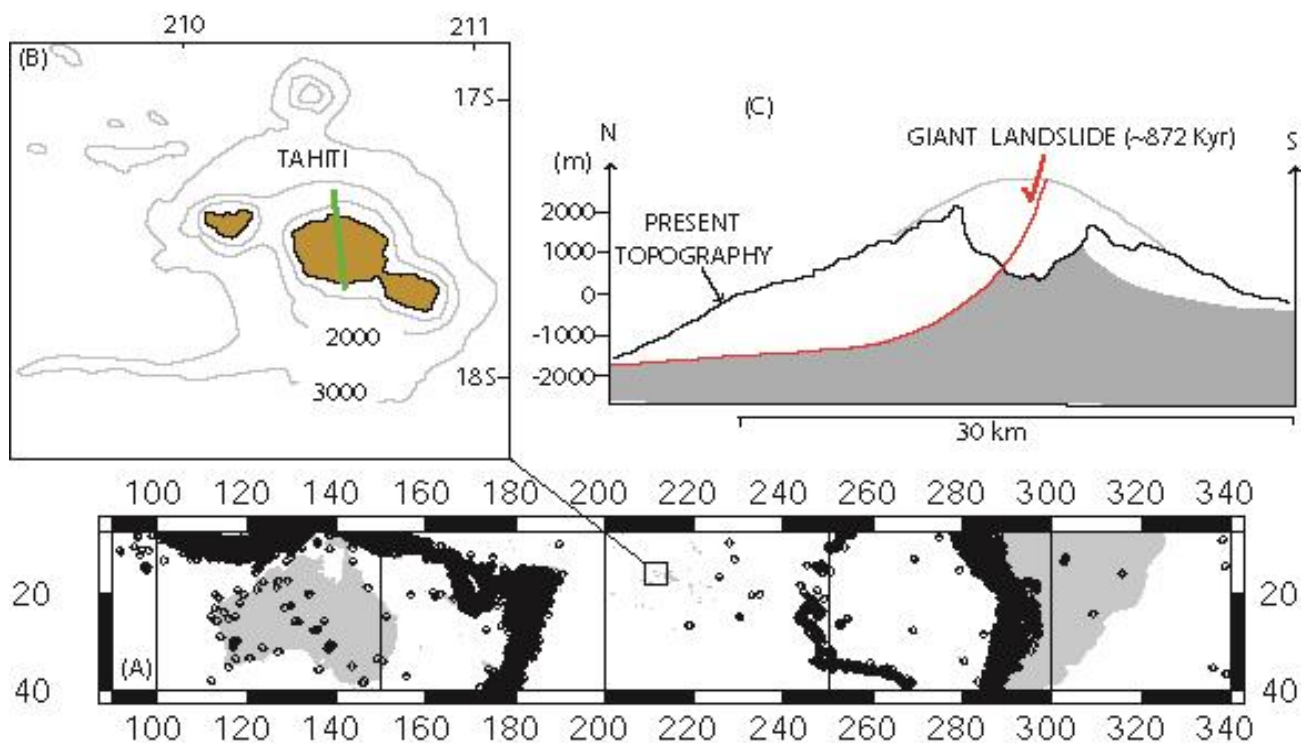

*Fig. 1: General context for the large Tahitian landslide. (A) Location of Tahiti (17.70° N, 149.40° E), circles represent the seismicity from 1997 to 2007 with a magnitude M>4 (seismological data from the USGS Earthquake Database). (B) Tahiti-Nui Island and location of cross-section, (C) Schematic cross-section of the giant landslide.*

## 2. Geological setting

The study area is located in the southern Pacific Ocean (Fig. 1A). Several submarine landslides occurred in French Polynesia (Clouard and Bonneville, 2004), whose archipelago includes the volcanic island of Tahiti. This is composed of two coalescent volcanoes, Tahiti-Nui to the northwest and Tahiti-Iti to the southeast (Clouard et al., 2001). Regionally, the Society volcanic chain consists of numerous volcanic islands and seamounts are aligned from northwest to southeast, with ages increasing to the northwest (Patriat et al., 2002). Tahiti-Nui (Fig. 1B) constitutes the youngest extinct volcano of the Society volcanic chain and is located near the Society hotspot (Duncan and McDougall, 1976).

Lateral collapses are visible on the northern and southern slopes of the main shield volcano (Clouard et al., 2001). The southern collapse occurred between 650 and 850 kyr (Clouard et al., 2001), while the northern collapse occurred around $0.872 \pm 0.01$ Ma ago (Hildenbrand et al., 2004). This large landslide may have been triggered by a paleoclimatic change (Quidelleur et al., 2008). Following the landslide, a second shield developed in the northern depression and dykes were generated on the depression's rims (Hildenbrand et al., 2004). Lava and breccias flowed into these depressions, filling them progressively (Hildenbrand et al., 2008). The present slope of the volcanic edifice ranges from 8–12°, similar to conditions before the large collapse (Hildenbrand et al., 2004 and 2006). The northern landslide involved volcanic rock several hundreds of metres thick, with a maximum deposit width of ~80–90 km. The maximum runout distance of the landslide front reached ~70–80 km, while the runout distance of the centre of mass was ~40 km (Fig. 2).

This volcano is located in an area where no significant seismicity has been recorded due to its intraplate location (Fig. 1A) and where tectonic deformation is negligible. Subsidence has been measured as 0.25 mm/yr (Bard et al., 1996) and 0.37 mm/yr (Fadil et al., 2009; Thomas et al., 2012). There have been no dyke intrusions or volcanic eruptions recorded during the past thousands of years, nor any uplift of the volcanic edifice due to magma reservoir activity.

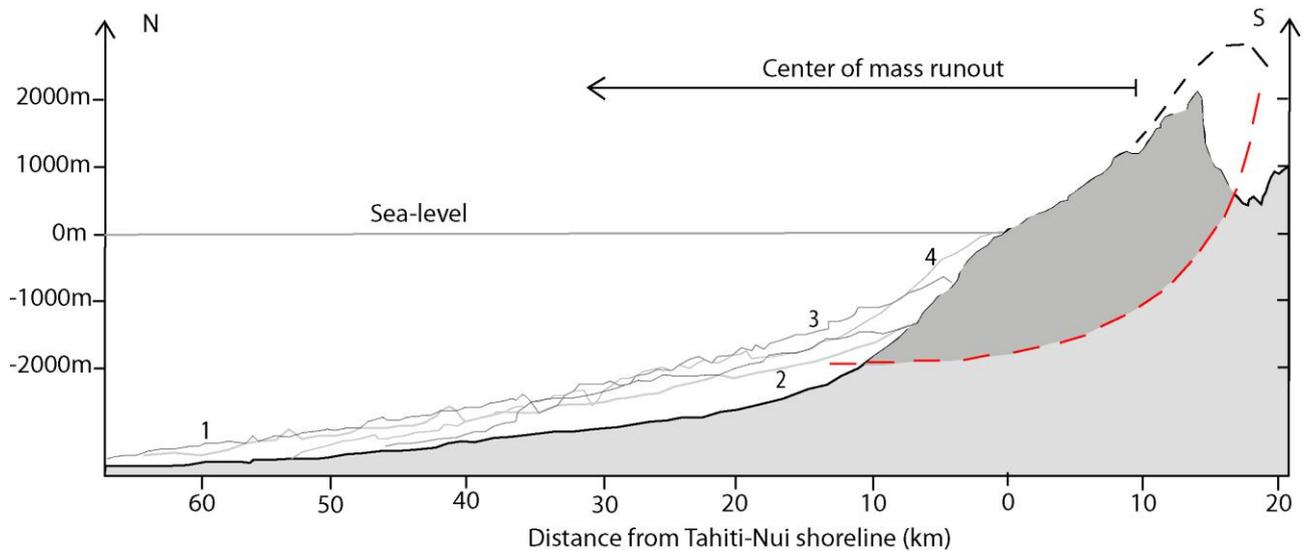

*Fig. 2: Longitudinal profiles along the large Tahitian landslide, from which the runout of the centre of mass was estimated as 40 ± 5 km (modified from Hildenbrand et al., 2008).*

## 3. Modelling approach

The mobility of the landslide was modelled in order to study various processes explaining the significant runout distance *L* and to assess the different event scenarios. The model analysed the propagation conditions for three cases: (i) completely subaerial, (ii) completely subaqueous, and (iii) subaerial initiation and subsequent subaqueous propagation.

The role of landslide volume on mobility has been investigated by several studies (Goren and Aharonov, 2007; Brunetti et al., 2009; Staron and Lajeunesse, 2009; De Blasio, 2011b), which showed that runout is proportional to material volume and that there is an apparent decrease of basal friction ($\mu \sim H/L$, where *H* is the height) in relation to increased rock volume. Nevertheless, even if complex thermo-poro-elastic mechanisms (lubrication by high pressure at the base, the presence of evaporites or ice, local melting, ground vibration, trapped air, etc.) are used to explain this phenomena, other studies have shown that slope path geometry and unconsolidated mass spreading play a significant role in describing mobility (Staron and Lajeunesse, 2009). In other words, mobility is controlled primarily by sliding processes in relation to classical friction laws and spreading processes in the case of large volume landslide as in other cases (Staron and Lajeunesse, 2009; Salmanidou et al., 2018).

Subaerial sliding along a slope $\theta$ was modelled using the weight $W$ of the sliding block and the frictional force $R$ at the base of the sliding block. To simulate the effects of a subaqueous environment, hydrostatic uplift due to buoyancy and the drag force $F_D$ due to seawater were also incorporated (Fig. 3). The latter was based on a classical fluid mechanics approach (Hürlimann et al., 2000):

$$F_D = 0.5 K \rho_f v^2 A \qquad (1)$$

where $K$ is a dimensionless shape coefficient for drag and skin friction, $\rho_f$ is the density of the fluid (i.e. water density if subaqueous or air density if subaerial), $v$ is the sliding block velocity and $A$ is the surface of the sliding block in contact with the fluid opposing the movement. The value of $K$ ranges from 0–2 (Hürlimann et al., 2000).

The equation of motion (Hürlimann et al., 2000) was defined as:

$$\frac{dv}{dt} = [g(\rho - \rho_f)(\sin\theta - \mu\cos\theta)/\rho] - [0.5(K/l)(\rho_f/\rho)v^2] \qquad (2)$$

where $t$ is time, $g$ is gravity, $\theta$ is the slope, $\mu$ is the coefficient of friction, $l$ is the length of the body, and $\rho$ is the density of the sliding block.

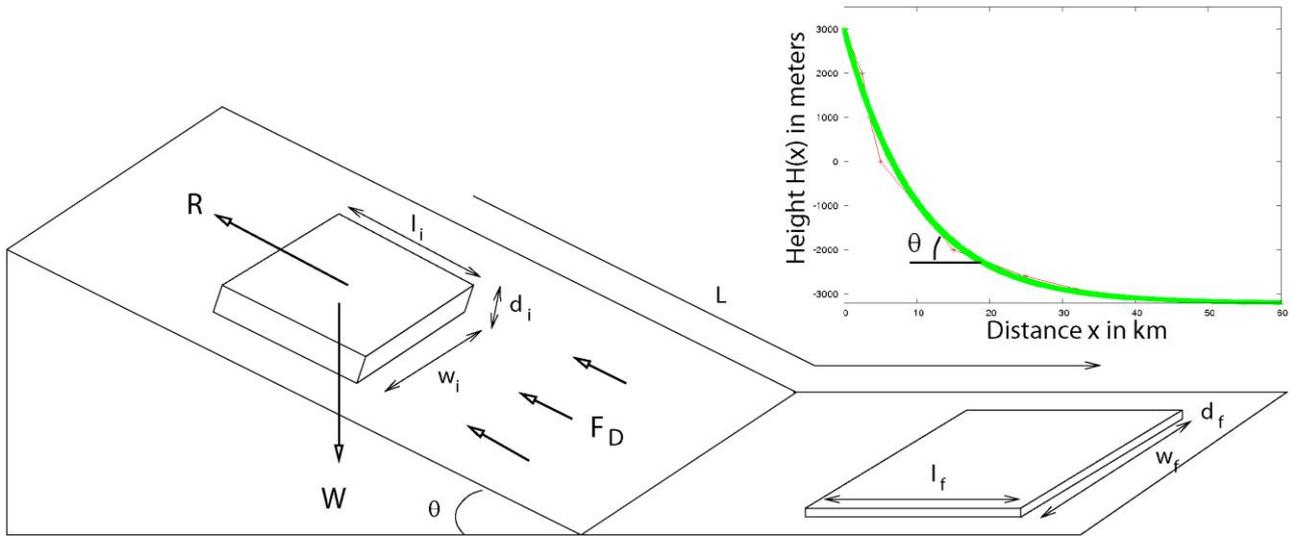

*Fig. 3: Geometrical parameters and forces applied to the landslide model along with the simulated path geometry, based on the equation $H(x) = H_0 \exp(x/l_0)$, where $H_0 = 6.2$ km and $l_0 = 10$ km.*

In the model, the transition from subaerial to subaqueous sliding was immediate without loss of energy. For simplicity, the slope path was represented by an exponential function, $H(x) = H_0\exp(x/l_0)$, where $H(x)$ is the height of the terrain and $x$ is distance along the path (Fig. 3), $H_0$ is the initial height, and $l_0$ is a length that allows to the curvature to fit the slide path (De Blasio, 2011a). These geometric parameters were set as $H_0 = 6.2$ km and $l_0 = 10$ km (Fig. 3). The density of the sliding block (composed of volcanic rocks) was considered constant at 2800 kg/m³. The density of the fluid $\rho_f$ was set as 1000 kg/m³ for subaqueous sliding and 0 kg/m³ for subaerial sliding. The geometry of the sliding body was characterised by an initial geometry (initial width $w_i$, initial length $l_i$, and initial thickness $d_i$), a final geometry (final width $w_f$, final length $l_f$, and final thickness $d_f$), and a sliding length $L$ (also referred to as runout) (Fig. 3). These parameters were estimated from geomorphologic and bathymetric data (Hildenbrand et al., 2004, 2006). The runout considered here was that of the centre of mass, not the maximum runout of the slide front.

The large landslide's initial conditions were estimated as width $w_i = 25$ km, length $l_i = 20$

km, and thickness $d_i$ = 2000 m, the final geometry was estimated as $w_f$ = 90 km, $l_f$ = 70 km, and $d_f$ = 160 m, and the sliding distance of the centre of mass was estimated as $L = 40 \pm 5$ km. The volume of the sliding block was identical at the initial and final states. The temporal evolution of the geometrical parameters from the initial state to the final state was assumed to be linear. Due to the short duration of the events considered (300 s < $t$ < 850 s), as defined by the simulations, this assumption did not significantly influence the maximum velocity and runout distance. Even if the real value was uncertain, the influence of the geometrical parameters was less significant (in the range of values tested) with respect to the mobility than the friction $\mu$ and the hydrodynamic factor $K$.

The parameters selected for the mobility analysis were the ratio between the hydrodynamic coefficient and the length of the sliding block ($K/l$) and the friction $\mu$. The first parameter is known to strongly influence sliding dynamics under subaqueous conditions (Hürlimann et al., 2000; De Blasio, 2011a). Assuming a parallelipedic shape of length $l$, width $w$, and thickness $d$, $K/l = C_D/l + C_S/d + 2C_S/w$, where $C_D$ is the drag coefficient and $C_S$ is the skin coefficient. The drag coefficient can be calculated using $C_D = 1.95 - 0.77 d/w$ (De Blasio, 2011a). The skin coefficient ranges from 0.002 < $C_S$ < 0.006 (De Blasio, 2011a). Under these assumptions, the parameter $K/l$ ranges from $5.10^{-5}$ m$^{-1}$ to $5.10^{-4}$ m$^{-1}$, similar to those used by Hürlimann et al. (2000) ($K/l = 2.10^{-4}$ m$^{-1}$ and $K/l = 2.10^{-3}$ m$^{-1}$). Under subaerial conditions, the coefficient $K/l = 0$ m$^{-1}$.

The value of friction $\mu$ can range from 0.6–0.85 (Bayerlee's law) for dry subaerial conditions of intact rocks and has been estimated as low as 0.0025–0.08 for subaqueous conditions in specific cases (Hürlimann et al., 2000; De Blasio, 2011a) such as hydroplaning or the presence of evaporites. In this study, the basal friction $\mu$ was set at an intermediate value of 0.2–0.3 under subaqueous conditions instead of 0.6 under subaerial conditions. The former value corresponded to an angle of friction of $\phi = 11-17°$ in accordance with the value considered appropriate for highly weathered and fragmented volcanic rock (Rodriguez-Losada et al., 2009).

Alternatively, a viscous rheology could be considered to simulate the complex behaviour of

the landslide using the Bingham law, in which the shear stress:

$\tau = \eta\, dv/dz + \tau_0$, if $\tau > \tau_0$

and $\tau = 0$, if $\tau < \tau_0$,

where $\eta$ is the viscosity, $\tau_0$ is the yield stress, $v$ is the velocity of the landslide, and $z$ is the vertical distance. For a thickness $h$ and a depth-averaged velocity $v$, the shear rate can be estimated by $dv/dz \approx 2v/h$ (Pelletier et al., 2008). The resulting shear stress thus becomes $\tau = 2\eta\, v/h + \tau_0$. Therefore, an increase in viscosity will increase the difficulty of landslide propagation, as expected. Under this assumption, landslide dynamics can be modelled by:

$$\frac{dv}{dt} = [g(\rho - \rho_f)\sin\theta\,/\,\rho] - [\,2\eta v\,/\,(\rho h^2) + \tau_0\,/\,(\rho h)\,] - [\,0.5\,(K/l)\,(\rho_f/\rho)\,v^2\,] \quad (3)$$

where the first term on the right side is due to the weight, the second term is due to the viscous behaviour of the landslide, and the third term is due to water drag.

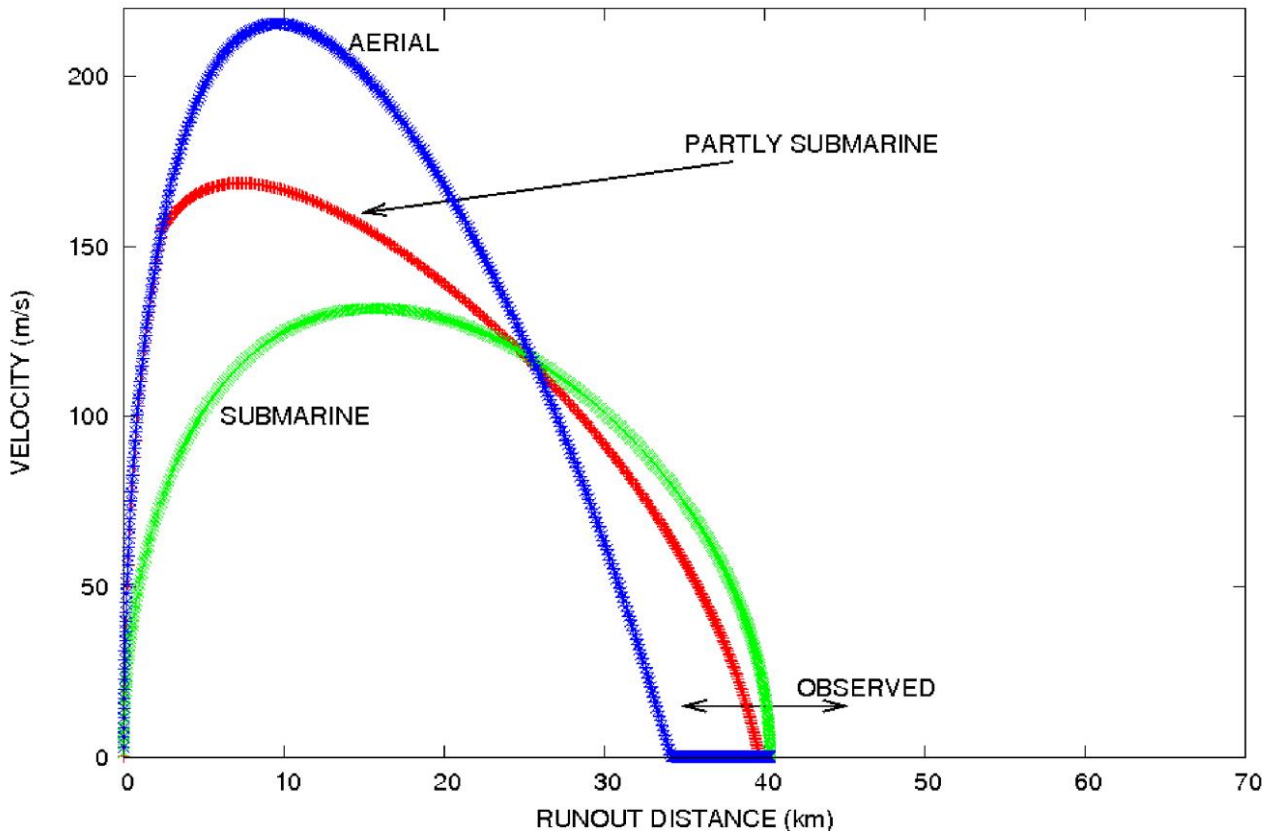

Fig. 4: Influence of three simulated propagation conditions: aerial, submarine, and partly

*submarine (5 km aerial, then submarine). $K/l = C_D/l + C_S/d + 2C_S/w$ for all cases. $\mu = 0.27$ ($\phi = 15.5°$) under submarine conditions and $\mu = 0.6$ under aerial conditions. $w_i = 25$ km, $l_i = 20$ km, and $d_i = 2$ km. Observed runout of the landslide's centre of mass is assumed to be $40 \pm 5$ km.*

## 4. Results

The landslide propagated mostly under subaqueous conditions although the slide may have initiated under subaerial conditions. This study's model assumed that the slide material entered into the sea after 5 km and compared the effects of subaqueous and subaerial propagation. The landslide's submarine and partly submarine mobility was of the same order of magnitude (Fig. 4). The runout distance and slide time could be higher under subaqueous conditions than under subaerial conditions, assuming a reduction of basal friction under submarine conditions from 0.6 to 0.3 (Fig. 4). The subaerial propagation of the landslide showed a higher maximum velocity than submarine propagation, despite the reduction of effective basal friction simulated in the latter (Fig. 4). When the effect of hydroplaning was not considered ($\mu$ was constant), aerial propagation was longer than submarine propagation due to the increased hydrodynamic drag force. The maximum aerial velocity was ~215 m/s, but only 170 m/s for a partly submarine landslide and 125 m/s for a submarine landslide (Fig. 4). Entering the water generally resulted in decreased velocity (Fig. 5). However, if the volume of material was significant ($L = 20$ km, $V = 1000$ km$^3$), the velocity of the landslide was able to continue increasing after entering the water (although at a smaller rate) for ~3 km due to the inertia of the sliding material (Fig. 6). The slide time was higher under subaqueous conditions than under subaerial conditions (Fig. 5).

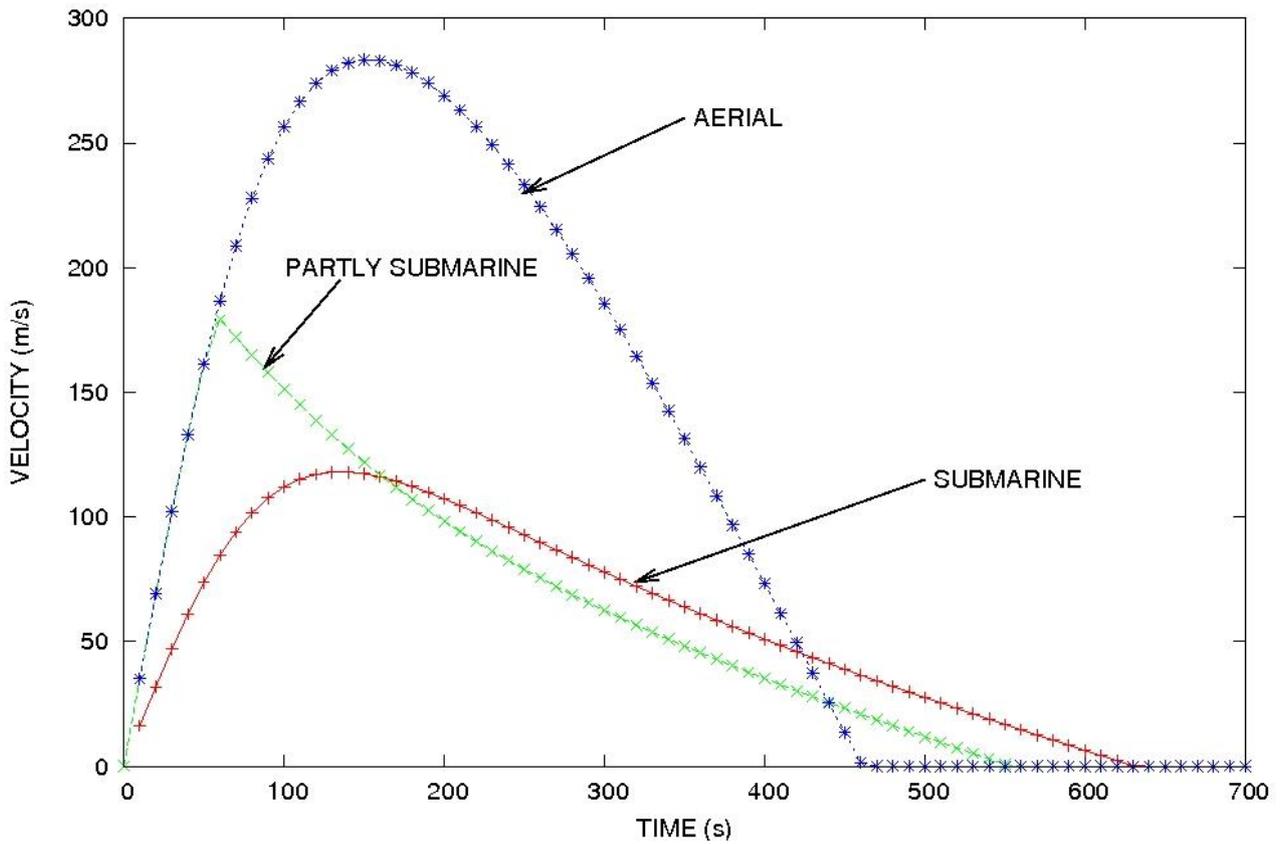

*Fig. 5: Landslide velocity over time under three different propagation conditions: aerial, submarine, and partly submarine (5 km aerial, then submarine). The drag friction was set equal to K/l = 0.0004. µ = 0.2 (ϕ = 11.3°) under submarine conditions and µ = 0.6 under aerial conditions. $w_i$ = 25 km, $l_i$ = 20 km, and $d_i$= 2 km.*

A landslide with a large initial volume $V$ of 1000 km$^3$ (volume $V = w$ x $l$ x $d$ = 25 km x 20 km x 2 km) allowed the runout distance of ~40 km observed offshore of Tahiti with a basal friction $\mu$ = 0.27 (Fig. 6A), but not with a basal friction $\mu$ = 0.2 (Fig 6B). Smaller landslides with a length $l$/10 or $l$/100 had a runout $L$ smaller than a bigger landslide with an initial length $l$ under partly submarine conditions. For a repetition of $n$ = 10 small landslides with a length $l$/10, a width $w$, and a thickness $d$ ($w$ x $l$/10 x $d$ = $V$/10), the total deposit volume ($V_{total}$ = 10 x $V$/10) was equivalent to a unique landslide with a volume $V$. Ten landslides with a volume $V$/10 had an expected runout of ~25 km when the basal friction $\mu$ = 0.27, below the observed runout distance observed (Fig. 6A).

However, when the basal friction $\mu = 0.2$, the runout distance of landslides with volume $V/10$ fit the observed runout data (Fig. 6B). For a volume of each landslide of $V/100$ (i.e. the length of each landslide became $\sim l/100$, but the initial width $w$ as well as the thickness $d$ was the same as for landslides with volumes $V$ and $V/10$), the calculated runout distance was $\sim$20 km for $\mu = 0.27$ and 30 km for $\mu = 0.20$, below the observed runout distance (Fig 6). For $l/100$, only a very small basal friction $\mu \ll 0.2$ was able to fit the observed runout of the centre of mass. The number of landslides with a volume $V/100$ necessary to obtain the observed volume of deposits was 100. In all simulations, the maximum velocity of the landslides ranged from 180–200 m/s.

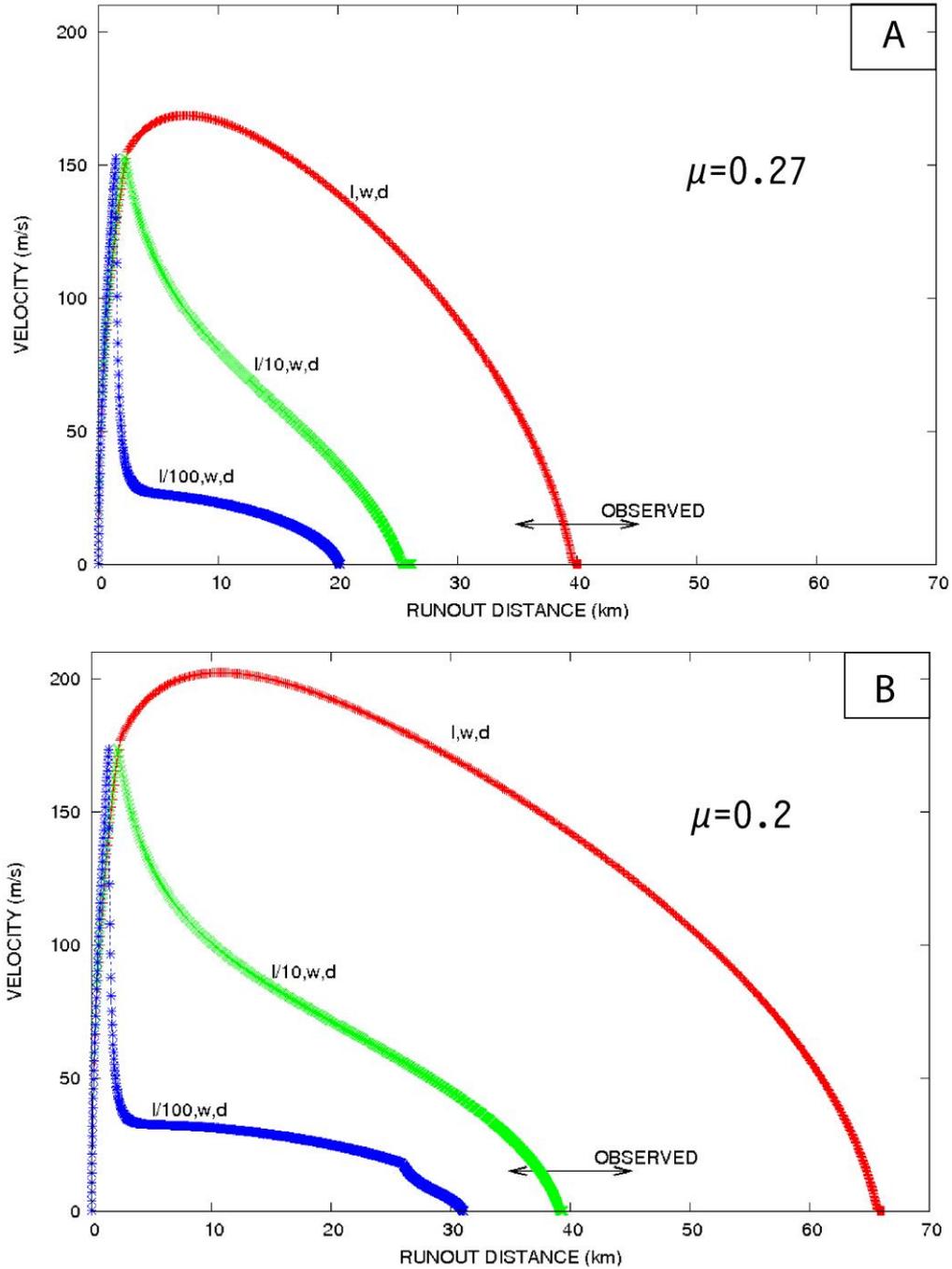

*Fig. 6: Effect of landslide volume on landslide dynamics for three volumes and two friction values: (A) $\mu=0.27$ ($\phi=15.5°$) under submarine conditions and (B) $\mu=0.2$ ($\phi=11.3°$) under submarine conditions. In both cases, propagation initiated under aerial condition but transitioned to submarine conditions after 5 km. The water effective drag coefficient was defined as $K/l=C_D/l+C_S/d+2C_S/w$. where $w_i=25$ km, $d_i=2$ km, and $l_i=20$ km, 2 km, or 0.2 km. The observed runout of the landslide centre of mass was $40\pm5$ km. Under aerial conditions, the basal friction $\mu=0.6$.*

## 5. Discussion

Several studies have discussed possible causes for low effective friction and high mobility observed at landslide fronts (Lucas and Mangeney, 2007; Goren and Aharonov, 2007; Staron and Lajeunesse, 2009; Johnson and Campbell, 2017; Crosta et al., 2018; Collins and Reid, 2020). In the Tahitian case considered here, the final deposit's centre of mass was located ~40 km from its initial position and the front of the slide propagated ~80 km. An effective basal friction $\mu$ of 0.2–0.3 under submarine conditions allowed the simulation to fit the observed runout distance. Reduced basal friction permitted the simulation of hydroplaning effects. Conversely, increasing aerial propagation increased the maximum velocity of the slide but slightly reduced the runout distance of the centre of mass.

Previous studies have suggested that landslide propagation under submarine conditions could result in longer runout distances and the occurrence of hydroplaning (Locat and Lee, 2002; De Blasio, 2011b; Hürlimann et al 2000). However, in several case where no hydroplaning occurred, submarine propagation reduced the runout distance (Gargani et al., 2014). Other lubrification effects could also explain decreases in basal friction under aerial or submarine conditions. The spreading of a large volume of material could partially explain the long apparent runout of the landslide's front (Johnson and Campbell, 2017).

This study's modelling showed that slide velocity decreased (Fig. 5) or reduced its acceleration (Fig. 4) after entering the water, as the drag force generated by water (Gargani, 2004c) is more efficient than that of air; the former could be significant at high velocities. Previous studies suggested that subaerial mass acceleration can be much greater than submarine mass acceleration yet (Kafle et al., 2016). Furthermore, the presence of water could generate a counterintuitive effect by increasing the runout distance (Fig. 4). Such significant runout under subaqueous conditions has been explained by hydroplaning (De Blasio, 2011a). The competition between drag and friction force on one hand, and gravity and buoyancy on the other, could explain the high runout under subaqueous conditions seen in this study. Under such conditions, the propagation along the path was slower but lasted longer (Fig. 5), as shown by previous studies (Hürlimann et al., 2000; De Blasio, 2011b). When the volume and length of a landslide are large, the inertia of the sliding block could allow a reduced effect of the drag force when entering the water, such that the sliding block speed is not reduced abruptly when the propagation becomes submarine (Fig. 6).

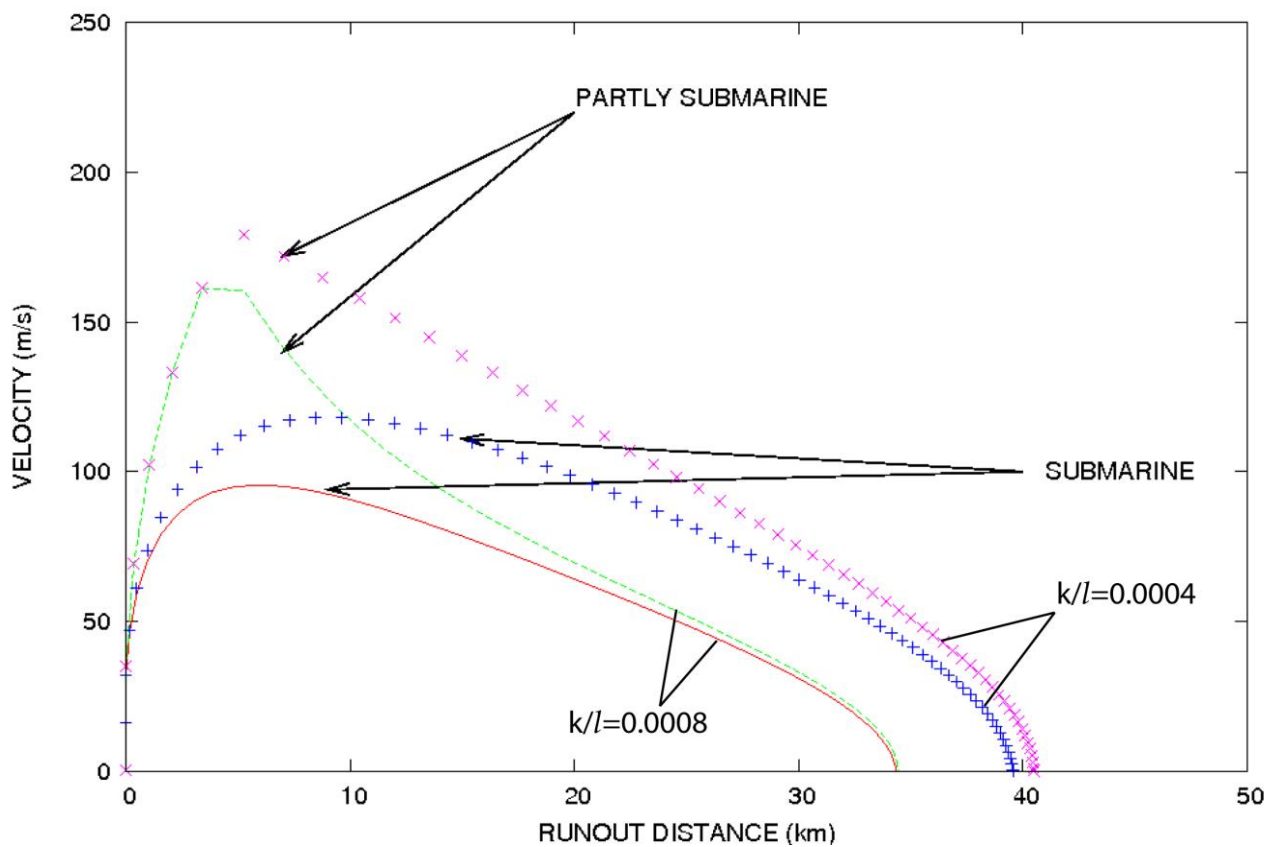

*Fig. 7: Influence of the effective drag coefficient k/l on landslide dynamics for submarine and partly submarine propagation. For aerial propagation, the effective drag coefficient k/l=0. µ=0.2 under submarine condition and µ=0.6 under aerial condition.*

One of this study's main questions relates to whether the Tahitian landslide deposits considered here were formed through a single slide event or multiple events. If the former, the runout distance of ~40 km could be easily achieved with a friction $\mu = 0.27$ under submarine conditions (Fig. 6A). However, under this condition, other landslide geometries did not fit the observed runout distance. The simulated runout was lower for smaller landslide geometries, as an increasing length $l$ decreases the ratio $K/l$. The drag coefficient is significantly reduced when the length of the sliding block is increased. Equation (2) shows that larger landslides are capable of propagating longer distances than smaller ones (Fig. 6). If all other parameters are equal, an increase in the effective drag coefficient $K/l$ will decrease propagation distance (Fig. 7). A decrease in $K/l$ by a factor of two can increase the runout of the centre of mass by a factor of ~1.2. In this model, the width of the landslide had only a small influence on the runout because the skin coefficient $C_S$ was small in comparison with the drag coefficient $C_D$.

Nevertheless, it may be possible that the deposits were formed by a limited number of landslide events (n<10). Indeed, the observed runout distance can also be explained by the occurrence of ten separate landslides with a length of ~2 km using a basal friction $\mu = 0.2$ under submarine conditions (Fig. 6B). However, this multi-slide scenario (n > 10) is less realistic, because it implies that the basal friction became very low ($\mu << 0.2, \phi < 10°$) in order to fit the observed runout. Although the occurrence of a single event was difficult to prove, the modelling showed that a reduced number of events (<10) was necessary to reproduce the observed runout distance. Therefore, these results support the proposal that a single event or a reduced number of events generated the large debris apron morphology. The ability of few events to form moderate to large landslide morphology (i.e. scars and mass transport deposits) has been advocated in several specific cases using more accurate geomorphological data and observations (Cauchon-Voyer et al., 2011 ; Puga-Bernabéu et al., 2017 ; Collins and Reid, 2020). For example, some studies of the Oso landslide have advocated for two larges, distinct failure events separated by a few minutes, whereas others studies suggested that the vast majority of landslide activity occurred within the initial minute (Collins and Reid, 2020). In comparison, the Gloria Knolls landslide consisted of three events (Puga-Bernabéu et al., 2017).

Extremely small basal friction values are not necessary to explain the large runout of the

Tahitian landslide's centre of mass, which primarily propagated under submarine conditions. Values of $0.2 < \mu < 0.3$ corresponded to an angle of internal friction from 11–17°, as expected for very weathered basalt or pyroclastic rocks (Rodriguez-Losada et al., 2009). To reach these values, the landslide may have caused significant fragmentation of the material. The numerous fractures and long polygenic history of the volcanic edifice probably contributed to the presence of weak material (Hildenbrand et al. 2004). The humid climate of Tahiti could also explain the highly weathered and altered nature of the rocks, which contributed to the presence of weak material from a mechanical perspective.

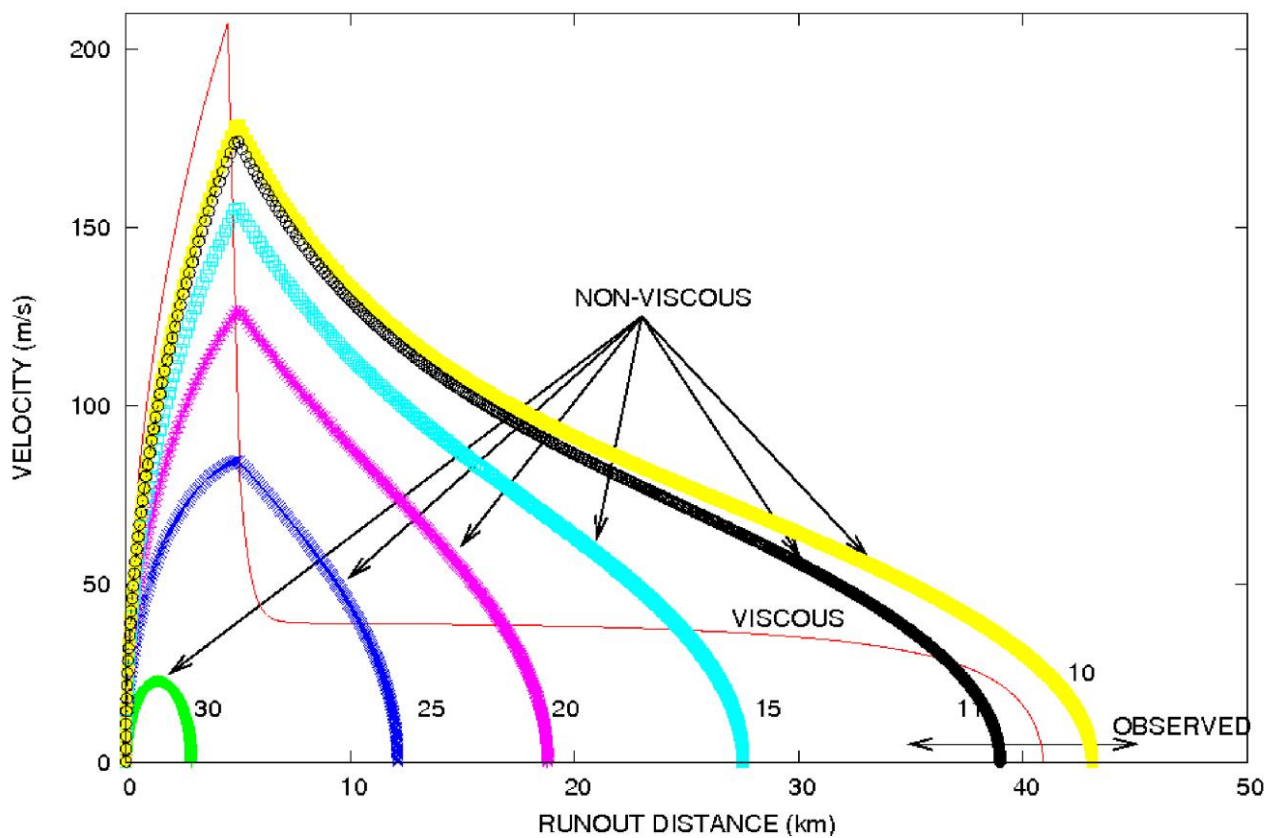

*Fig. 8: Modelled landslide velocity vs. runout distance ($\eta = 5$ MPa.s, $\tau_0 = 0.5$ Mpa) for viscous and various non-viscous rheologies for different angles of internal friction ($\phi = 30°, 25°, 20°, 15°, 11°,$ and $10°$). In all cases, the landslide propagates aerially for the first 5 km, then transitions to submarine conditions. $K/l = C_D/l + C_S/d + 2C_S/w$. $w_i = 25$ km, $l_i = 2$ km, $d_i = 2$ km. The observed runout of the landslide centre of mass is $40 \pm 5$ km.*

The estimated slide peak velocity ranged from 125–250 m/s, within the range of other large landslide peak velocities observed or calculated in volcanic contexts (Voight et al., 1983; Hürlimann et al., 2000) but higher than that of smaller submarine landslides (20–50 m/s) (L'heureux et al., 2013; Rodriguez et al., 2017; Abril and Periañez, 2017; Salmanidoue et al., 2018; Sawyer et al., 2019). The significant mass displacement may have triggered an isostatic rebound (Gargani, 2004, 2004b, 2010). A tsunami should also have occurred after the landslide as has been estimated for other comparable events (Bohannon and Gardner, 2004; Rodriguez et al., 2017). Erosion processes (Gargani et al., 2006; Cojan et al., 2007; Gargani et al., 2016) should have occurred after the landslide event.

The mechanical behaviour of the highly fragmented sliding material was accurately simulated by low effective friction and allowed explanation of its significant mobility. Alternative modelling experiments including more complex rock behaviour may also explain the dynamics of this landslide (Fig. 8). For example, viscous behaviour could also explain the observed runout of the centre of mass (Fig. 8). There is an abrupt decrease of the velocity from the aerial part to the submarine part. In this case, the submarine velocity was < 50 m/s for almost all slide paths (Fig. 8) and smaller than for non-viscous conditions; this is near the peak velocity obtained for other submarine landslides with a smaller size (L'heureux et al., 2013; Rodriguez et al., 2017; Abril and Periañez, 2017; Salmanidoue et al., 2018; Sawyer et al., 2019).

## 6. Conclusion

The velocity of the Tahitian landslide exceeded 125 m/s, allowing the fragmentation of highly weathered rocks. The effective basal friction for submarine propagation was relatively low ($0.2 < \mu < 0.3$), enabling significant mobility for the centre of mass (~40 km$^3$) and suggesting that hydroplaning occurred under submarine conditions. The long runout length of the centre of mass also suggested that the landslide deposit was the consequence of a single event or a reduced number of events (n<10). The initial length of the landslide varied between $l/10 < l_i < l$, where $l = 20$ km for a total volume $V \sim 1000$ km$^3$. The significant mobility of the material and the slide dynamics constrained the geometry of the mass transport deposit.

**Declarations of interest**: None

**Funding** : This research did not receive any specific grant from funding agencies in the public, commercial, or not-for-profit sectors.

**Acknowledgements:**